# Human-AI Co-Learning for Data-Driven AI

Yi-Ching Huang[1,2], Yu-Ting Cheng[1,2], Lin-Lin Chen[2], Jane Yung-jen Hsu[3],
[1]National Taiwan University of Science and Technology
[2]Eindhoven University of Technology
[3]National Taiwan University
{y.huang3, y.cheng, l.chen}@tue.nl, yjhsu@csie.ntu.edu.tw

**ABSTRACT**

Human and AI are increasingly interacting and collaborating to accomplish various complex tasks in the context of diverse application domains (e.g., healthcare, transportation, and creative design). Two dynamic, learning entities (AI and human) have distinct mental model, expertise, and ability; such fundamental difference/mismatch offers opportunities for bringing new perspectives to achieve better results. However, this mismatch can cause unexpected failure and result in serious consequences. While recent research has paid much attention to enhancing interpretability or explainability to allow machine to explain how it makes a decision for supporting humans, this research argues that there is urging the need for both human and AI should develop specific, corresponding ability to interact and collaborate with each other to form a human-AI team to accomplish superior results. This research introduces a conceptual framework called "Co-Learning," in which people can learn with/from and grow with AI partners over time. We characterize three key concepts of co-learning: "mutual understanding," "mutual benefits," and "mutual growth" for facilitating human-AI collaboration on complex problem solving. We will present proof-of-concepts to investigate whether and how our approach can help human-AI team to understand and benefit each other, and ultimately improve productivity and creativity on creative problem domains. The insights will contribute to the design of Human-AI collaboration.

**INTRODUCTION**

Machine learning and AI have been widely employed in a variety of applications to help people to make decisions on high-stakes application domains, ranging from healthcare and criminal justice decision making to semi-autonomous driving. Humans and AI as a team has great potential to solve complex problems because they have distinct mental models and complementary capabilities that can be combined to augment each other towards superior results. Such fundamental differences offer opportunities to bring multiple perspectives to reach better outcomes; however, it can also cause unexpected failure, resulting in harmful experience and serious consequences. Therefore, there is an urging need to explore ways for facilitating collaboration between humans and AI.

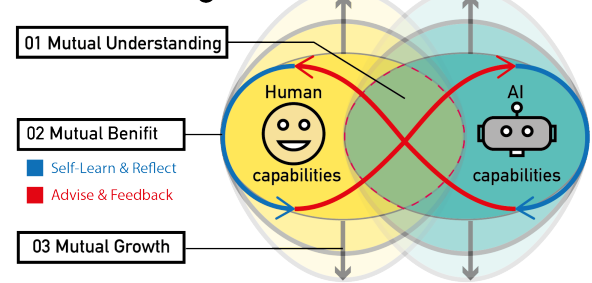

Figure 1. Three goals of co-learning are to support humans and AI (1) building mutual understanding, (2) facilitating mutual benefits, and (3) enabling mutual growth over time.

Many researchers have increasingly interests in facilitating human-AI collaboration in complex problem solving [6, 4]. Most work mainly focuses on enhancing intelligibility or explainability of AI systems that allow machines to explain how they make decisions and why they fail to people [2, 11, 14, 3]. Some studies devise approaches to mitigate or recovery AI errors to achieve users' expectations and enhance user acceptance of machines' suggestions [7, 5]. However, human and AI are dynamic, learning entities; that is, their mental models and capabilities are changing over time. Rather than enhancing the explainability of AI systems, people need to learn how to interact and how to teach or train AI systems to fulfill their expectations [10, 8]. Moreover, we need a complete perspective to consider human-AI collaboration rather than focusing on either machine or human perspectives. That is, we should consider both sides as a whole to explore what kind of characteristics of the human-AI team should have and how to facilitate the human-AI team to evolve together to adapt to such a dynamic and hybrid situation.

In this research, we argue that humans and AI systems should learn from/with each other, grow over time. Instead of correcting AI errors, we should both leverage or discover AI's specific advantage to create a win-win situation and develop a better 'us.' Therefore, this research extends the notion of Human-Centered Machine Learning and proposes a conceptual framework called "Co-learning," in which human or AI in a team is able to interact and learn from/with, and grow with their collaborator over time. Three key characteristics are required for facilitating Human-AI collaboration in the context of creative domains: "mutual understanding", "mutual reciprocity", and "mutual growth". We discuss the challenge of human-AI collaboration and why co-learning matters for enabling a human-AI team to accomplish superior results in the context of creative application domains.

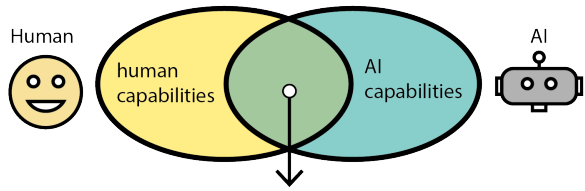

Figure 2. Humans and AI build mutual understanding through an interactive process.

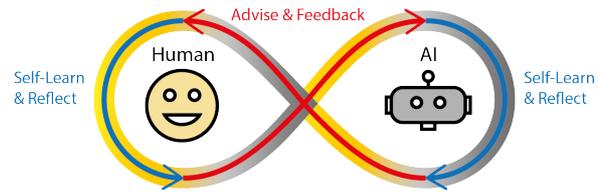

Figure 3. Human and AI establish a positive loop by continuous feedback and adaption.

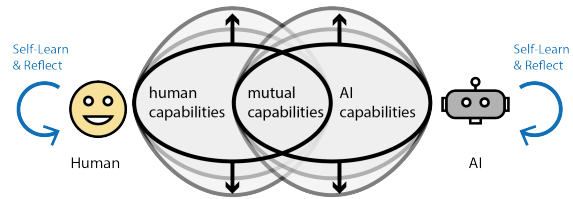

Figure 4. Human and AI are growing and expanding their capabilities over time.

In this work, we will show proof-of-concepts to demonstrate whether or how co-learning can facilitate designer/researchers to make sense of data, build new insights with their AI partners in the context of creative problem domains. First, we explore whether of how AI systems support people to understand the relationship between data and AI algorithms and take full advantage of AI-specific ability to solve the collective problems. We attempt to design an AI playground which designers can play with data, annotations, and several standard algorithms to build understanding of AI. During this play, designers will learn how to clean data, label data, select a proper algorithm based on the data they have, and teach their AI to acquire specific ability to help them detect objects and trace movement of objects on image data. Second, designers and their AI partners will tackle problems together. In this process, human and ai will build mutual understanding, facilitate mutual benefits, and enable mutual growth for humans and AI, and ultimately result in better productivity and creativity. A series of studies will be conducted to evaluate our proposed framework on creative task domains. The findings will contribute to either HCI or AI community for designing Human-AI collaboration for creative application domains.

## CO-LEARNING FOR IMPROVED HUMAN-AI TEAM

### Definition of Co-Learning

Co-learning indicate that human or AI in a team has "the ability that can interact and learn from/with, and grow with their collaborator." The goal of co-learning is to support two "dynamic growing entities" to "build mutual understanding, facilitate mutual benefit, and enable mutual growth" over time (see Figure 1). In the notion of co-learning, AI need to learn how to explain what it thinks, how it behaves, and why it makes a decision to humans; on the other hand, humans also need to learn how to represent human's intention to AIs, and explore ways of teaching AI, and adapt vocabulary to allow AI to learn appropriately.

With co-learning, human and AI can form a team to achieve superior results than ever before successfully. To achieve this goal, there are three key requirements: mutual understanding, mutual reciprocity, and mutual growth. We briefly discuss these concepts below.

### Key Concepts of Co-Learning

*Mutual Understanding*
Human and AI have different mental models, capabilities, and behaviors. To facilitate human-AI collaboration, the fundamental step is to develop a mutual understanding (i.e., shared mental model) between humans and AI. We define mutual understanding as the ability of learning entities (Human or AI) to expect others and to be expected by others (see Figure 2). As with human-human collaboration, the effectiveness of human-AI collaboration is rooted in a shared understanding of each collaborator's capabilities [1]. In specific, they need to identify not only strengths but also weaknesses through an iterative and interactive process because both of them are growing and changing over time.

*Mutual Benefits*
Built on the mutual understanding, human and AI need to establish reciprocity to mutual benefit each other by complementing or augmenting each capability with a goal of achieving superior results that people or AI cannot achieve alone (see Figure 3). They can help each other by identifying or correcting mistakes, learning from unexpected situations, and discover new possibilities. It can also amplify each ability to become a better team. Over time, a positive feedback loop can facilitate trust-building between humans and AI.

*Mutual Growth*
In this team, humans and AI both have "growth mindsets" (co-evolving); that is, they learn together, learn from each other, learn with each other, and evolve/grow over time. In specific, both of them can "self-reflect" on their own based on other inputs or inner understanding, and then self-regulate their learning strategies to develop new or revise their ability or capability (see Figure 4). The self-learning loop is to update two learning entities' mental models to adapt to the changing environment.

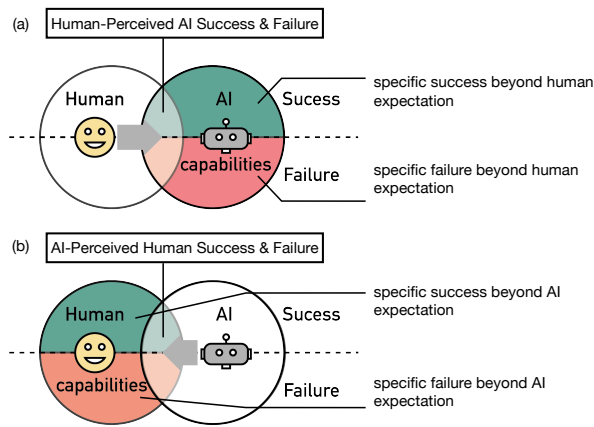

**Figure 5.** (a) Human-perceived AI success and failure (top image). (b) AI-perceived human success and failure (bottom image).

**How Co-Learning Facilitates Human-AI Collaboration**

*Reducing mismatch between Human and AI expectations over time*: Human and AI are two different growing entities. First, human and AI have different mental models, abilities, and vocabulary. To work together, the first step is to build a mutual understanding and identify fundamental differences or mismatch between how humans and AI perceive, reason about the world, and how they react, perform, and interact with the world. Second, human and AI are continuous learning over time. Their perceptive ability and mental model also change over time, resulting in a dynamic mismatch. Therefore, co-learning aims to develop a growing understanding for both of them and adapt to the changing nature.

*Becoming better through human-AI complement and augmentation*: Human and AI both make mistakes. Recent research suggests that AI needs to explain their decision-making process to humans, especially for AI-specific failures or errors [14]. Such types of errors are only made by AI and beyond human expectations, leading to unexpected and harmful consequences (e.g., semi-autonomous car accidents and misdiagnoses by clinical decision support systems). Instead of understanding AI failures, humans also need to learn how to correct machine failure or verify machine predictions by providing data or annotations [10]. While a human is able to correct AI errors or verify AI outputs, AI is able to become "better" than ever before. On the other hand, people and AI also have different types of success (see Figure 5). In order to take full advantage of AI, people also need to understand the capability of AI and what is their specific advance, allowing them to make a better plan to incorporate AI into their decision process.

*Building trust through continuous feedback and adaptation*: People usually cannot trust AI systems because of their uncertainty, biases, and failure [13]. Recent researchers have growing interests in exploring ways of building trust between humans and AI [9, 2]. While people without any technical background involved in the process of building or training AI/ML models, they are more likely to trust machine predictions than professional AI experts [12]. Therefore, we argue that people co-learn with their AI partners over time, and then humans and AI can develop a trust relationship. With a trust relationship, people are more receptive to the suggestions from their AI partners, and AI updates its model effectively by prioritizing the inputs given by its human partners.